\documentclass[showpacs,apl,twocolumn,floatfix]{revtex4}

\usepackage{graphicx,float,amsmath}
\usepackage{mathrsfs}
\usepackage{natbib}

\begin{document}
\date{\today}

\title{Charge and spin transport  over record distances  in  GaAs metallic n-type nanowires : II nonlinear charge transport.}

\author{H. Hijazi$^1$}
\author{D.~Paget$^2$}
\author{G. Monier$^1$}
\author{G. Gr\'egoire$^1$}
\author{J. Leymarie$^1$}
\author{E. Gil$^1$}
\author{F. Cadiz$^2$}
\author{C. Robert-Goumet$^1$}
\author{Y. Andr\'e$^1$}

\affiliation{%
$^1$ Université Clermont Auvergne, CNRS, SIGMA Clermont, Institut Pascal, F-63000 Clermont-Ferrand, France}

\affiliation{%
$^2$Physique de la mati\`ere condens\'ee, Ecole Polytechnique, CNRS,  IP Paris, 91128 Palaiseau, France}

%\affiliation{%
%$^3$Ioffe Institute, Saint Petersburg, Russia.}

%\affiliation{%
%$^4$ McMaster University Canada. Hamilton Canada}

\begin{abstract}
We have investigated the photocarrier charge  transport in n-type metallic GaAs nanowires ($\approx  10^{17}$ cm$^{-3}$ doping level), grown by hydride vapor phase epitaxy (HVPE) on Si(111) substrates. Analysis of the luminescence intensity spatial profiles for selected energies in the spectrum allows us to determine the spatial distribution of photoelectrons, minority photoholes and electrons of the Fermi sea as a function of distance from the light excitation spot.  This analysis shows that charge can be transported over record  distances larger than $25$ $\mu$m at 6K, in  spite of the expected localization of minority holes in the potential fluctuations generated by statistical fluctuations of the donor concentration. It is shown that  transport is little affected by the fluctuations because of the build up of large internal electric fields which strongly increase the hole and electron mobilities and therefore enable transport. Comparison of the spatial profiles of the emissions due to hot electrons and to the Fermi sea gives evidence for at least three spatial zones, including a zone of excess intrinsic electrons near the excitation spot and a zone of depletion of these electrons at a distance larger than 2-10 $\mu$m depending on excitation power. The internal outward electric field increases the kinetic energy of photoholes in the fluctuations  so that after a given distance to the excitation spot, there occurs ballistic transport over the fluctuations.     
\end{abstract}
\maketitle

\section{Introduction}
%\label{intro}

In a companion paper,  hereafter referred to as [I], it has been shown that n-type NWs near the nonmetal/metal Mott transition are most promising   for spin transport. Photoelectron transport in the Fermi sea leads to a rapid establishment of charge equilibrium between the two reservoirs. However, the spin reservoirs remain distinct  and spin order can be transported over distances as large as 20 $\mu$m.\ 

An immediate consequence of the above finding is that, since spin is carried by photoelectron charge, photocarrier  charge is transported over similar distances. However, for such NWs, it may seem  difficult to achieve minority hole transport over long distances. Indeed, because of the efficient relaxation of the hole kinetic energy \cite{chebira92, bimberg1985} and  because of the difficulty of holes to undergo tunnel processes due to their large effective mass \cite{levanyuk1981}, holes may stay trapped in the potential fluctuations. These fluctuations are related to the tails in the conduction and valence bands, caused by  the statistical spatial fluctuations of donor concentration \cite{efros1972, shklovskii1984, lowney1986, borghs1989}. In these systems, investigations have considered luminescence analysis of bulk systems \cite{levanyuk1981}, as well as  transport of majority carriers on bulk materials, using macroscopic tools such as conductivity measurements \cite{benzaquen1987, cutler1969, anderson1958}.\ 

The present work is an experimental investigation of charge transport in conduction and valence band tails of n-type GaAs NWs already used in [I]. These NWs have a donor doping level $N_D$ in the low  $ 10^{17}$ cm$^{-3}$ range  \cite{hijazi2019} that is, about one order of magnitude larger than the one of the Mott transition. Using a CW microluminescence technique, we monitor  the luminescence intensity  spatial profiles as a function of distance to the excitation spot. This is performed at 6K and over distances as large as  25 $\mu$m along the NW. From a comparison of the results as a function of  energy in the spectrum, it is found that  two key effects enable minority carrier transport over long distances. Firstly, there occurs a significant spatial redistribution of intrinsic electrons, producing large macroscopic  electric fields. Secondly, these electric fields strongly increase the  hopping probability and therefore the mobility of photocarriers.\ 

In the same way as for  electrolyte cells with which the present system has close analogies\cite{chazalviel1990}, several phases  appear in the spatial profiles, which depend on excitation power. These phases are caused by the appearance of an excess of charge in the Fermi sea near the excitation spot, along with a converse depletion at a distance, depending on excitation power, between 2 and 10 $\mu$m. \

This paper is organized as follows. The following section is devoted to the experimental aspects.  In Sec. III, we present the charge  spatial profiles at the corresponding energies and for several excitation powers. The spatial distributions of internal electric field and mobile charges are characterized in  Sec. IV and Sec. V contains a qualitative explanation of the nature of the various spatial phases in the intensity profiles.

%While resolution of the drift-diffusion equation for one dimensional transport in a NW leads to  a featureless exponential spatial decay of the photoelectron and hole concentrations,

\section{Experimental}
\label{exp}
%\subsection{NW growth and preparation}
All the results presented here were obtained on the same NW as in [I]. This gold-catalyzed NW, HVPE-grown on Si(111) substrates  has a length of $\approx 80$ $\mu$m and a diameter of $\approx 220$ nm. The  donor doping level $N_D$ is in the  low  $ 10^{17}$ cm$^{-3}$ range \cite{hijazi2019}.  This NW was treated immediately after growth by a nitrogen plasma, in order to reduce the surface oxidation under air exposure and the surface recombination velocity \cite{mehdi2019}. Finally, the NW was deposited horizontally on a grid of lattice spacing 15 $\mu$m.\
 
%\subsection{Experimental}
  
The experimental setup is the same as in [I]. The excitation light is a tightly-focused, continuous-wave laser beam (Gaussian radius $ \approx 0.6\; \mu$m, energy $1.59$ eV). The luminescence light is focused on the entrance slit of a spectrometer equipped with a CCD camera as a detector. In the present work, one considers sections of the CCD images along a direction parallel to the entrance slit. Such sections give the intensity  spatial profiles at a given energy in the spectrum. Since analyis of weak PL signals at long distance from the excitation spot requires an improved dynamic range, the laser light was carefully removed by a filter and background light was rejected by comparing results with the laser on the NW with results with the laser slightly displaced out of the NW. Under these conditions, the spatial profiles could be investigated up to distances as large as  $25$ $\mu$m from the excitation spot.\

\begin{figure}[tbp]
\includegraphics[clip,width=8.6 cm] {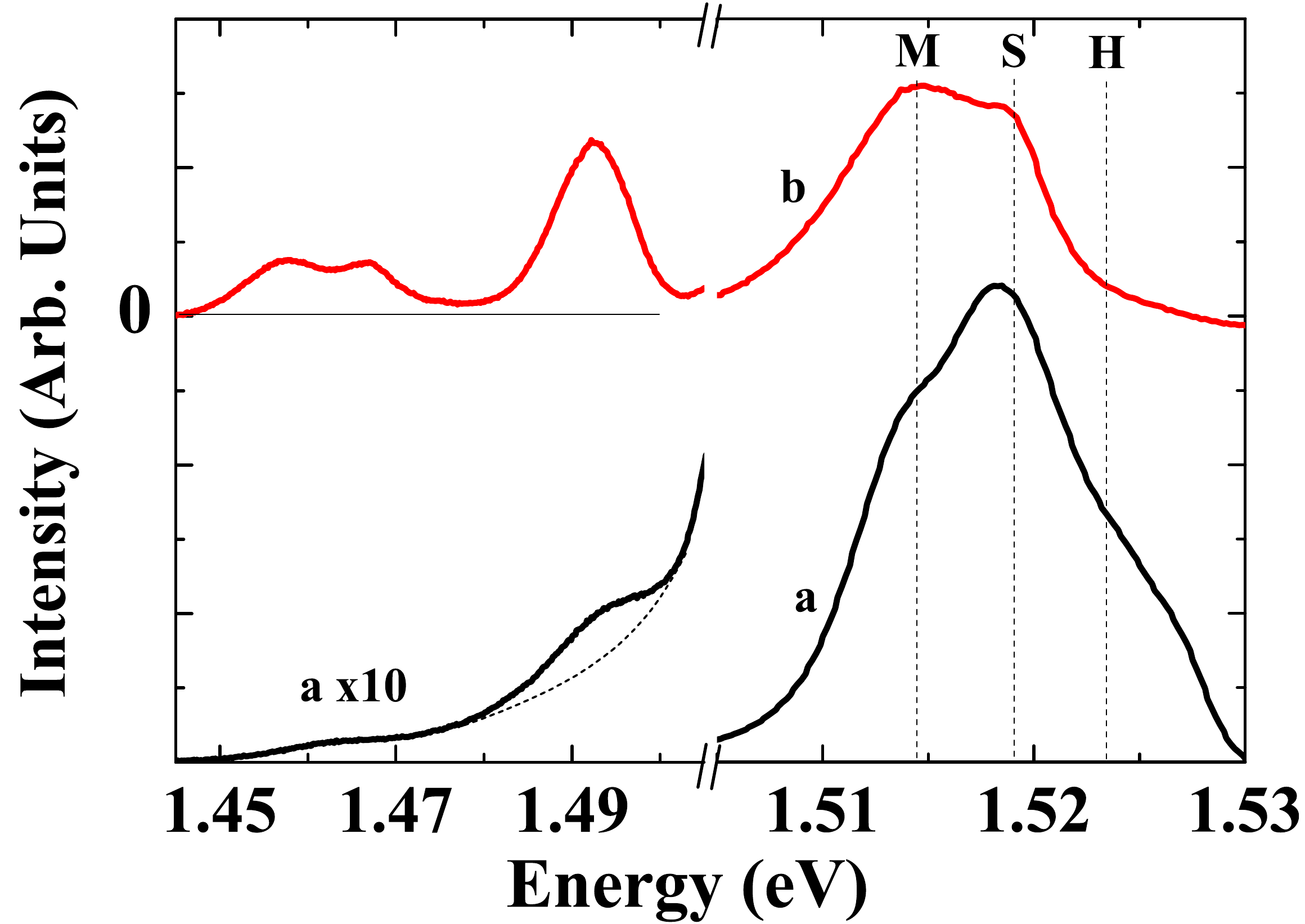}
\caption{Intensity spectra, taken at the excitation spot, for an excitation power of 1 mW (Curve a), and for a very small excitation power of  $9$ $\mu$W (Curve b). These spectra exhibit the nearbandgap luminescence composed of line M near 1.515 eV, of a shoulder labelled S and a hot electron feature, labelled H.  Also shown in Curve b  are  the acceptor-related signals at 1.493 and 1.46 eV, respectively. These signals are no longer resolved at high excitation power (Curve a) because of the low-energy tail of the near bandgap luminescence. }
\label{Fig1}
\end{figure}

Fig. \ref{Fig1} shows, for specificity, spectra taken at the excitation spot, obtained by section of the image along an axis perpendicular to both the enrance slit and the NW. Curve b  shows the intensity spectrum  at $z=0$ for an excitation power of  $9$ $\mu$W, corresponding  to a very small effective power density of $\approx 30$ $W/cm^2$. The spectrum is composed of i) the  band at 1.493 eV, due to residual shallow acceptors, probably  carbon \cite{kisker1983, skromme1984}, ii) of a composite signal composed of a LO phonon replica of the acceptor emission near 1.457 eV and a line near 1.463 eV. This line is attributed to shallow acceptors states perturbed by nitrogen impurities\cite{leymarie1989,ruhstorfer2020}. iii) Of the near bandgap emission, which  is itself composed of  a main line near 1.515 eV, labelled M, a shoulder at 1.159 eV (S) and  a high-energy tail  above 1.525 eV (H).\ 

It has been shown in [I] that line M  is due to recombination of minority photoholes with intrinsic electrons. Its intensity at a given radial position $r$ and axial position $z$, is given by 
\begin{equation}\label{Imain}
	 I_{main}= K n_0p. 
	\end{equation} 	
where $K$ is the bimolecular recombination constant. Here,  $n_0$ is the intrinsic electron concentration and $p$ is the photohole concentration. Lines S and H are due to recombination of Fermi edge photoelectrons and of hot photoelectrons, respectively, with the same minority photoholes as line M. The corresponding  emission intensities are   
\begin{equation}\label{Ihot}
	 I_{hot}= K_{hot} np. 
	\end{equation}	 	
\noindent
where $n$ is the photoelectron concentration and the bimolecular recombination constant $K_{hot}$ is likely to be different from  $K$.\ 
 
Note finally that, at  high excitation power (Curve a), in the same way as for p-type material \cite{cadiz2017} the acceptor emissions are relatively weak and there appears a   featureless tail which extends down to 1.44 eV. This tail is a manifestation of the  density of states  tails known to be present in the conduction and valence bands due to disorder. The corresponding emission is due to recombination between photoelectrons localized in zones of positive  potential and photoholes localized in zones of negative potential and the luminescence energy is decreased by the internal electric field.\

\section{Energy-resolved spatial intensity profiles}

\begin{figure}[tbp]
\includegraphics[clip,width=8.6 cm] {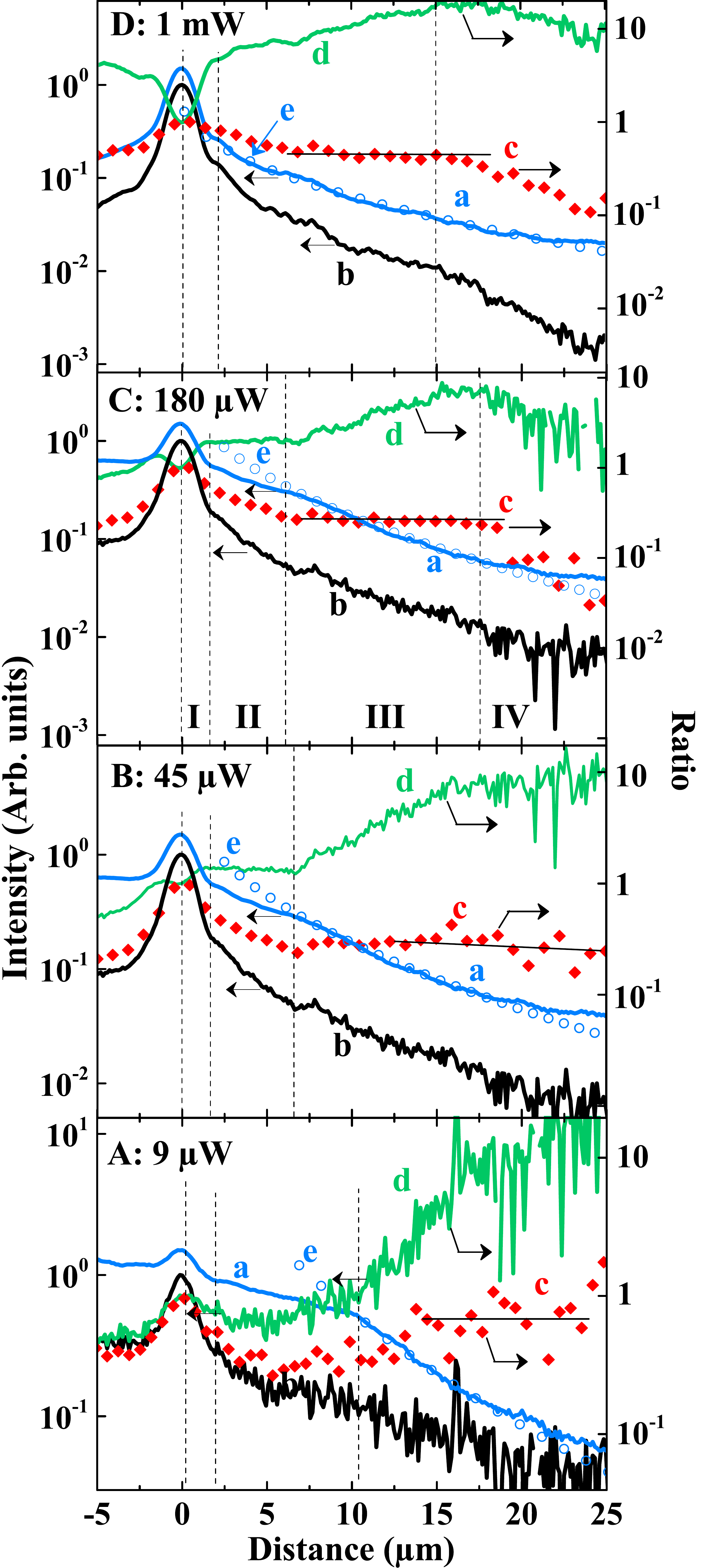}
\caption{Intensity spatial profiles  of line M  (1.515 eV, Curves a), of line H ( 1.527 eV, Curves b) for selected excitaton  powers of  9 $\mu$W (A),  45 $\mu$W (B) 180 $\mu$W (C) and 1 mW (D). Each panel also shows the ratios $\mathscr{C} $  (Curves c) and $\mathscr{B}$ (Curves d), given by Eq. \ref{B} and Eq. \ref{C}, respectively. Curves e are fits of Curve a  using Eq. \ref{ballistic}, describing ballistic hole transport over the fluctuations.}
\label{Fig2}
\end{figure}

Fig. \ref{Fig2} shows, for selected  excitation powers, the spatial intensity profiles at 1.515 eV (Curves a, line M) and at  1.529 eV (Curves b, line H). The spatial profile of line S, due to photoelectrons out of thermodynamic equilibrium with the  Fermi sea, has been considered in [I] and will not be discussed here. Four distinct regimes, labelled I-IV in Panel C, can be distinguished in the profiles shown in Fig. \ref{Fig2}. For line M  at the smallest  excitation power (Curve a of Panel A), the profile exhibits  a  relatively weak decrease up to about $2$ $\mu$m (Phase I) and a  slow  exponential-like, decay up to  $11$ $\mu$m  from which we obtain an effective diffusion length of  $10.8$ $\mu$m (Phase II) and again a faster one for larger distances (Phase III). The profile of line H  does not exhibit the slow decay of Phase II. It consists mostly of a  decay up to  2 $\mu$m, for which  the amplitude  is larger than that of Curves a, followed by a single exponential-like decay over the whole spatial range.\
Increase of the excitation power only slightly affects the profile of line H, apart from the  appearance of a faster decay beyond $20$ $\mu$m (Phase IV, most visible in Panel D of Fig. \ref{Fig2} ). On the other hand, the power increase  progressively induces two changes in the profile of line M : i) The amplitude of phase I  strongly increases. ii) The slow decay of the intensity profile in phase II occurs over a progressively smaller distance and becomes faster. For the largest excitation power (Panel D), Phase II has completely disappeared and the profile consists of only phases I,  III and IV.\

 \begin{figure}[tbp]
\includegraphics[clip,width=8.6 cm] {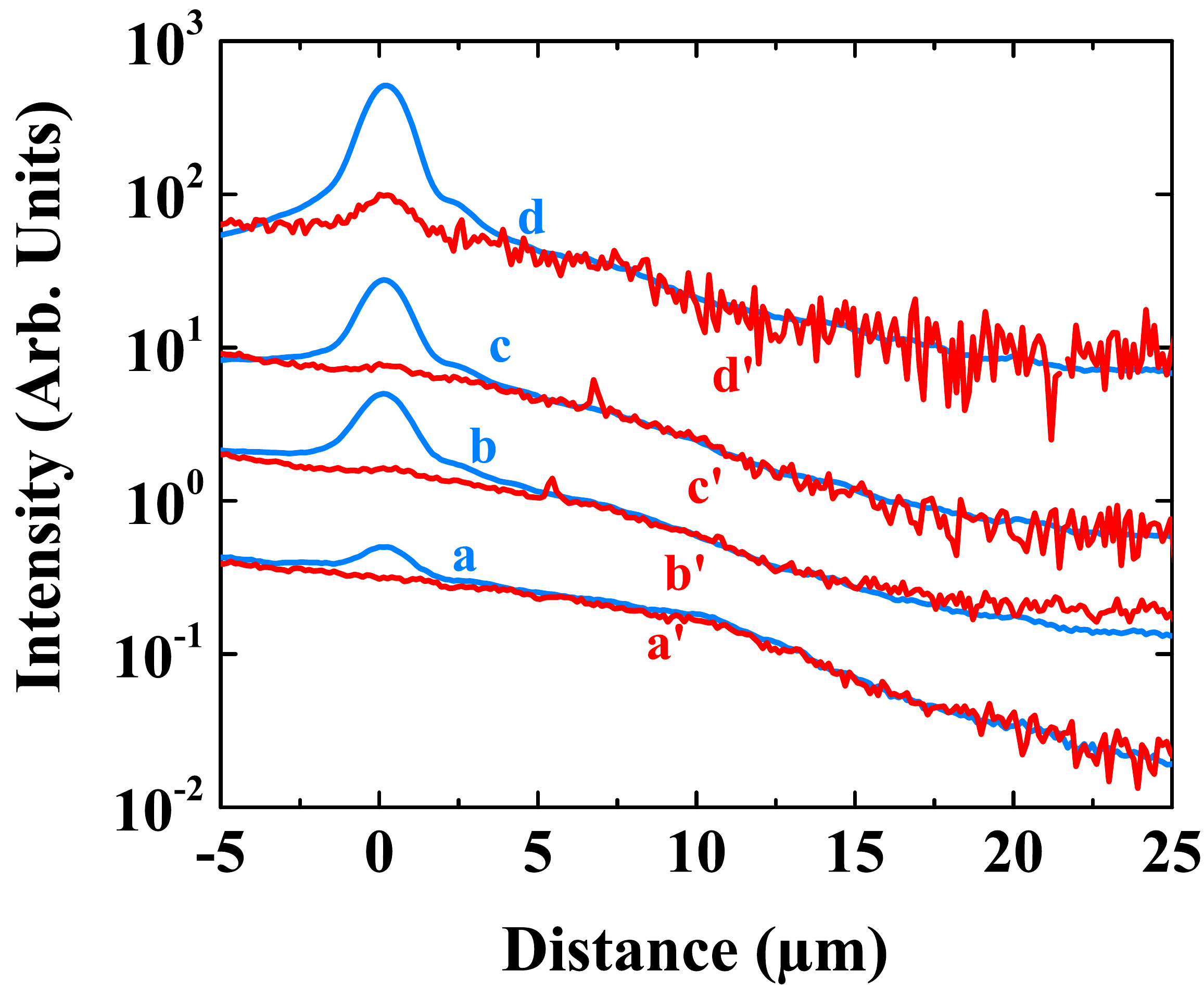}
\caption{ Luminescence intensity spatial profiles for the acceptor emission at 1.454 eV and for line M, for an excitation power of 9 $\mu$W (Curve a and a'),  45 $\mu$W (Curve b and b'), 180 $\mu$W (Curve c and c') and 1 mW (Curve d and d'). The profiles were shifted vertically for clarity and  are remarkably  similar  with the noteworthy exception that the  acceptor profiles do not exhibit the rapid decay near the excitation spot.}
\label{Fig3}
\end{figure}

Fig. \ref{Fig3} shows the spatial profiles of the acceptor emission at 1.454 eV. This energy is chosen in order to minimize the perturbation induced by the low-energy tail of line M at high excitation power. Apart from the rapid decay near the excitation spot, which is not observed on the acceptor emission, these profiles  coincide nearly perfectly with those of line M. This correspondence is natural, since the two emissions occur through recombination of holes (localized in the fluctuations of the valence band or trapped at acceptors, respectively) with the same intrinsic electrons. The absence of a rapid decay of the acceptor emission near the excitation spot is interpreted as a saturation of the acceptor levels, most of whom are neutral near the excitation spot since the  hole concentration is larger than that of residual acceptors.

\section{Interpretation}
 
The purpose of the present section is to explain  why, as seen from  observation of intensity profiles up to 25 $\mu$m, minority photoholes can be transported over very large distances in spite of the potential fluctuations of the valence and conduction bands. These results are in  contradiction with the fact that the relaxation of the hole kinetic energy \cite{chebira92, bimberg1985} is more rapid than for electrons, so that holes tend to accumulate in the  potential wells, of height larger than the thermal energy,  where tunneling processes are less probable than for electrons because of their large effective mass.\

We can exclude a simple explanation implying photon-mediated transport, originating from laser or luminescence light channeling in the NW. Indeed, no light at the laser energy is found at the end of the NW, which is evidence that the excitation laser does not couple to guided modes in the NW. Some luminescence light  may propagate along the NW and may be  reabsorbed over a distance of the order of 1 $\mu$m. It could affect the spatial profile  over longer distances if the  newly-generated electron-hole pairs in turn emit photons (photon recycling). However, because of the  matrix elements involved in these processes, emission of a photon in the same way as  creation of a spin-polarized electron, occur with a loss of angular momentum by a factor of 2,  so that spin recycling occurs with a loss of a factor of 4 and should create weakly spin-polarized electrons, in contradiction with the results of [I].\

Another effect which is to be excluded is the possible  presence of thermoelectric charge and spin currents due to spatial inhomogeneities of the photoelectron temperature. The temperature spatial profiles, reported  in [I], show a very weak temperature spatial gradient at low excitation power. At the highest power,  the gradients are very close to those reported before in similar experimental conditions \cite{cadiz2014}. Since the latter work has concluded that thermoelectric currents do not strongly contribute to the profiles, thermoelectric effects will be neglected here.\    

Thus, the intensity spatial profiles directly  reflect photocarrier transport along the NW. As shown in Ref. \cite{baranovskii2006}, even  for hopping transport, it is possible to define effective mobilities and diffusion constants.  Charge  transport in the NW is described by two conservation equations. The first one is the diffusion equation for minority holes. 

\begin{equation}\label{eqdiftrous}
	 g  - Kn_0 p    - K_{hot}np -  \frac{p}{\tau _{nr}} +\vec{\nabla }[D_{h}\vec{\nabla } p + \mu_h p \vec{E}]= 0. 
	\end{equation} 

\noindent
Here $g$ is the rate of creation of electron-hole pairs, $\tau _{nr}$ is the nonradiative recombination time, $D_{h}$ is the hole diffusion constant, $\mu_h$ is the hole mobility and $\vec{E}$ is the internal electric field, arising from distinct transport properties of electrons and holes. The  electron diffusion equation is 

\begin{align}
\label{eqdifchar}
	 g  - Kn_0 p  &  - K_{hot}np -  \frac{n}{\tau _{nr}} + \nonumber \\ & \vec{\nabla }[D_{e}\vec{\nabla } (n+  \delta  n_0) + \mu_e (n+ n_0^*) \vec{E}]= 0.  	  
 \end{align}

\noindent
where  $D_{e}$ is the diffusion constant and  $\mu_e$ is the electron mobility. Here  $\delta n_0 = n_0 -N_D$ is the light-induced change of the concentration  of intrinsic electrons and  $ n_0 ^* k_B T_e / E_{Fe} $  is  the reduced concentration  of electrons in the Fermi sea which can participate to drift currents.  Here $q$ is the absolute value of the electron charge,  $k_B$ is Boltzmann's constant and $E_{Fe}$ is the electron Fermi energy. If the internal electric field  $\vec{E}$ is negligible in Eq. \ref{eqdiftrous}, hole diffusion is unipolar and is decoupled from electron diffusion.  Resolution of the resulting equation  leads  in the present case to a featureless exponential decay of photocarrier concentration (see supplementary material), at variance with the experimental results.

\subsection{Internal electric fields} 

Within the usual ambipolar diffusion scheme, it is known that electrostatic interaction between mobile negative and positive charges \cite{smith1978, cadiz2015c, cadiz2015d, cadiz2017b} can lead to nonexponential intensity spatial profiles \cite{paget2012}. Comparison of Eq. \ref{eqdiftrous} and Eq. \ref{eqdifchar} gives the ambipolar electric field 
 
\begin{equation}\label{champel}
	\vec{E_a} = \frac{(D_h-D_e)\vec{\nabla} p } {\mu_e (n+ n_0^*)+\mu_hp}  +\frac{D_e \vec{\nabla} (n+ \delta n_0 -p)} {\mu_e (n+ n_0^*)+\mu_hp}
	\end{equation} 
\noindent
This equation  allows us  to write Eq. \ref{eqdiftrous} as a drift-diffusion equation with an  ambipolar diffusion constant given by  
 \begin{equation}    
D_a= \frac{ \mu_e (n+ n_0^*)D_h +\mu_h p D_e}{\mu_e (n+ n_0^*)+ \mu_h p} 
 \label{Da} 
 \end{equation}  

\noindent 
The effective diffusion length is $L_a = \sqrt{D_a \tau_h}$, where the hole lifetime $\tau_h$ is given by  $1/\tau_h = K_{hot} n+K n_0 + 1/\tau_{nr}$.\

In the present case these equations can be simplified for two distinct reasons. Firstly,  
although minute departures from neutrality are possible, caused by  internal electric fields, it has been shown that charge neutrality is valid  \cite{cadiz2015c}, so  that   
\begin{equation}    
p   \approx  \delta n_0  + n    
 \label{neutr} 
 \end{equation}
 
\noindent 
This implies that the second term of  Eq. \ref{champel} can be neglected.  Secondly,  taking a typical value of the diffusion constant  of $100$  cm$^2/s$ \cite{cadiz2014}, the photocarrier concentration $n$ at the excitation spot is of the order of $10^{14}$ cm $ ^{-3}$ for the weakest excitation power. The fact that the result is several orders of magnitude smaller than the doping level and the fact that possible departures from the monomolecular regime at the highest excitation power are not observed allow us to conclude that $n <<n_0^*< N_D$ and $p<<n_0^*< N_D$ throughout the excitation power range. In this case  $D_a \approx D_h$ and $L_a = \sqrt{D_h \tau_h}$ so that unipolar diffusion takes place. As a result, the electric field can be approximated by 

\begin{equation}\label{champel2}
	\vec{E_a} \approx \frac{(D_h-D_e)\vec{\nabla} p} {\mu_e n_0^* +\mu_hp}  
	\end{equation} 
\noindent 
   
This equation expresses the fact that the difference of hole and electron diffusive currents is equal to the difference  of the corresponding drift currents. Assuming $D_e > D_h$, this field is directed outwards and proportionnal to the slope $\vec{\nabla} p$ of the intensity profile of line M.  The field $E_a$ given by Eq. \ref{champel2} is evaluated using the Einstein relation, which is for a disordered sample $D_e=\mu_e \mathscr{E}/q$. The energy $\mathscr{E}$ is comparable to the fluctuation amplitude at low temperature and becomes equal to the usual value $k_BT_e$, if the temperature is increased \cite{baranovskii2006}.  One obtains $E \approx \mathscr{E}p/(qL_a n_0^*)$. This field is of the order of several $10^{-4} (p/n_0^*)$ $V/\mu$m.  This is several orders of magnitude smaller than  typical electric fields in the fluctuations, of the order of the unscreened effective field near a donor  $E_D/a_0^* \approx 0.6$ $V/\mu$m. As a result, usual ambipolar fields cannot affect photocarrier transport in the NW and therefore cannot explain the spatial profiles. 

It is now shown that, as already shown earlier experimentally \cite{wolfe1973},  the presence of disorder may strongly increase the mobility and therefore the length over which carriers can be transported. Here, the potential fluctuations, rather than preventing transport, will result in a  self-adjustment of the  electric field to a value  enabling significant transport.  The reason is the strong dependence of the conductivity $\sigma$  on electric field  similar to that observed for amorphous materials \cite{baranovskii2006}, implying an increase of carrier mobility with the electric field. In order to obtain the electric-field dependence of the mobility, one recalls first the temperature dependence of the electron and hole  conductivities, of the type  $\sigma_{e(h)} = \sigma_{e(h)0} \exp\left[ -  \left( \Delta_{e(h)}/k_BT_{e(h)}) \right)^{\beta} \right] $. Here, $\Delta_{e(h)}$ are activation energies and  the exponent $\beta$ is of the order of unity  \cite{benzaquen1987, baranovskii2006}.\

 \begin{figure}[tbp]
\includegraphics[clip,width=8.6 cm] {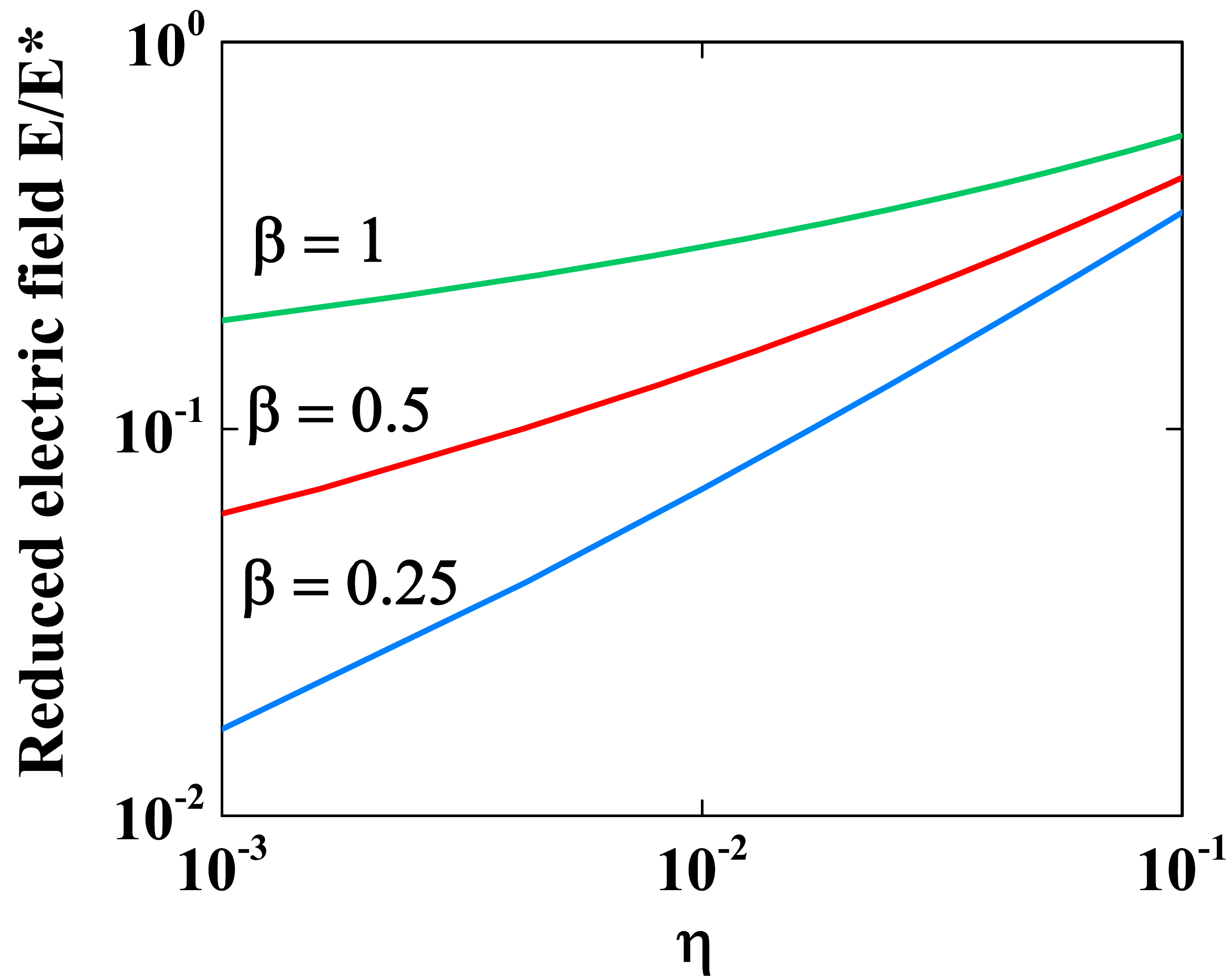}
\caption{Dependence of the reduced electric field as a function of $\eta$ given by  Eq. \ref{eqnonlin}, for an  exponent $\beta$, given by the thermal dependence of the conductivity, equal to 0.25 (insulating phase), 0.50 (metallic phase close to the transition) and 1. } 
\label{Fig4}
\end{figure}

Since thermal and electrical activation of the conductivity are of the same nature, the electric field dependence of the mobility is found using the above equation by replacing the thermal energy by $qE\delta$, where  $\delta$ is the characteristic  distance travelled in an elementary hopping process \cite{baranovskii2006}. This distance can be larger than the typical dimension of the fluctuation, of the order of the donor Bohr radius, if variable range hopping processes are significant. As a result,  the electron and hole mobilities must be expressed as \

\begin{equation}    
\mu_{e(h)}(E)= \mu_{e(h)}^* \exp\left[-\left(\frac{\Delta _{e(h)}}{q E  \delta}\right)^{\beta}  \right],   
 \label{mob} 
 \end{equation}
 
\noindent
where   $\mu_{e(h)}^*$ are the mobilities at large electric fields.  Since the effect of electric field on carrier transport is expected to be weaker for the less localized electrons than for the holes, one expects   $\Delta _{h }>\Delta _{e }$. The electric field is obtained using Eq. \ref{champel2} and Eq. \ref{mob} and is the solution of a nonlinear equation. In the present case where the hole drift current is smaller than the electron one ($ \mu_h^*p < \mu_e^*n_0^*$), the current balance is weakly affected by the hole drift current, so that the reduced electric field $u=E/E^*$, where $E^*= \Delta_e/(q\delta)$, is the solution of     
$u \exp\left[-\left( 1/u \right)^{\beta}  \right] =\eta$, where 
\begin{equation}    
\eta  =  \frac{( D_h -D_e)\nabla p  } {\mu_e^*n_0^* E^* }  \approx  \frac{p}{n_0^*} \frac{\mathscr{E}}{\Delta_e} \frac{\delta}{L_a}    
 \label{eqnonlin} 
 \end{equation} 
is the ratio of diffusive currrents to the drift current of the Fermi sea in the field $E^*$. The approximate expression, found using Einstein's relation, shows that $\eta$ is a fraction of unity. The dependence of the reduced electric field as a function of $\eta$  for several values of $\beta$ is shown in Fig. \ref{Fig4}. For $\beta =1/4$ which is appropriate  for the insulating phase  \cite{benzaquen1987},  the reduced  electric field increases from  $10^{-2} $ to  $ 10^{-1}$  for $  \eta $ between several $10^{-4}$ and several $10^{-2}$. Near the transition ($N_D \approx 10^{16}$ cm$^{-3}$), one has $\beta \approx 0.5$. The electric field becomes larger and its dependence on $\eta$ becomes weaker. For the present doping level, values of $\beta$ have not been reported. It is natural to expect a further increase of $\beta$, for which, as seen in Fig. \ref{Fig4} for $\beta=1$, $E$ is a significant fraction of $E^*$ in a wide range of values of $\eta$. As a result, there occurs a self-adjustment of the electric field and therefore of the hole drift current according to Eq. \ref{mob}.  The spatial redistribution of mobile charges which produces this field is investigated in the following subsection. \  

\subsection{Charge redistribution.} 

\begin{figure}[tbp]
\includegraphics[clip,width=8.6 cm] {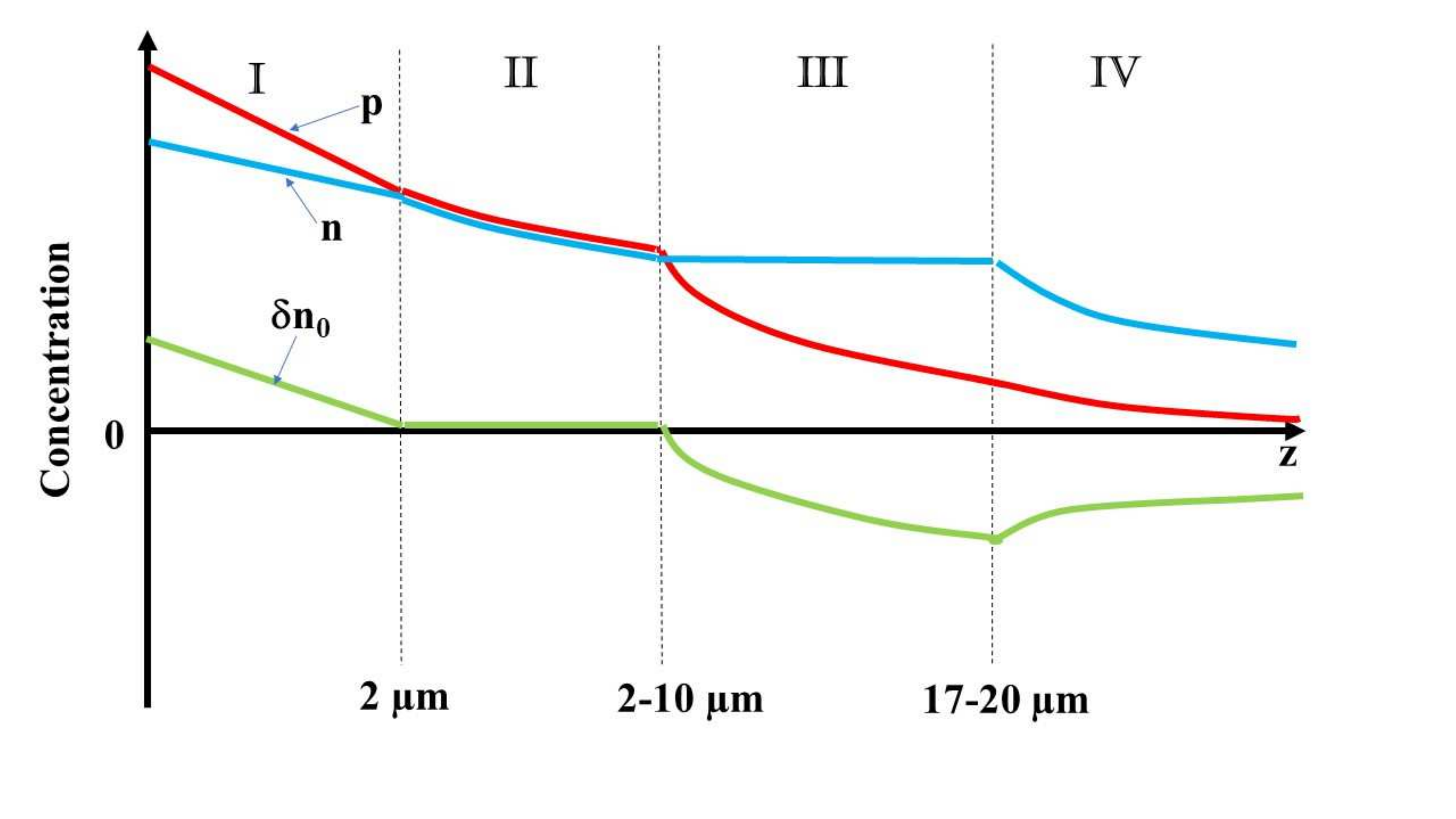}
\caption{Illustration of the spatial dependences of concentrations of photoelectrons $n$, photoholes $p$ and of the relative concentration of intrinsic electrons $\delta n_0$ (see Sec. IV. A), for the various spatial phases defined in  Fig. \ref{Fig2}. The limits of these spatial phases depend on excitation power and their range of values is indicated. The amplitude of the Fermi sea redistribution $\delta n_0$ is negligible for an excitation power of $9$ $\mu$W and increases with excitation power. }
\label{Fig5}
\end{figure}

Calculation of the charge spatial profiles should be  performed using a resolution of Eq. \ref{eqdiftrous} and Eq. \ref{eqdifchar},  using the modified values  of mobilities and  electric field. These equations are highly-coupled nonlinear equations and their resolution is beyond the scope of this work. Here, the charge redistribution of the photocarriers and of the Fermi sea is determined independently on models describing transport, from the spatial profiles of  Fig. \ref{Fig2}.  The results, illustrated in  Fig. \ref{Fig5}, show that there is a depletion of intrinsic electrons at large distance from the excitation spot, compensated by an excess near the excitation spot. The various phases in the spatial profiles are directly related with this spatial redistribution and are characterized  by distinct electric field  and  carrier mobility values.\

In order to determine the spatial profiles of the three terms of Eq. \ref{neutr} separately, one defines  the  quantity  

%\begin{align}
%\label{B}
%  \mathscr{B} =\frac{I_{hot}}{I_{main} ^2} \approx \mathscr{B}_0 \frac{n}{p}  \approx   %\mathscr{B}_0 (1-\frac{\delta n_0}{p}). 
% \end{align}

\begin{align}
\label{B}
  \mathscr{B}= \frac{I_{hot}}{I_{main} ^2} & =  \mathscr{B}_0 \frac{1-\delta n_0/p }{(1+\delta n_0/N_D)^2} \nonumber \\ & \approx   \mathscr{B}_0 (1-\frac{\delta n_0}{p}) =  \mathscr{B}_0 \frac{n}{p}. 
 \end{align}

%\begin{equation}    
%\mathscr{B}= \frac{I_{hot}}{I_{main} ^2} = \mathscr{B}_0 \frac{1-\delta n_0/p }{(1+\delta n_0/%n_0)^2} \approx   \mathscr{B}_0 (1-\frac{\delta n_0}{p}).    
% \label{Bbis} 
% \end{equation}

\noindent
where $I_{main}$  and $I_{hot}$ are given by Eq. \ref{Imain} and Eq. \ref{Ihot}, respectively, $ \mathscr{B}_0= K_{hot}/[( K N_D) ^2] $ and the exact expression of  $\mathscr{B}$ in this equation is obtained using  Eq. \ref{neutr}.  In the same way, one defines 
\begin{equation}    
\mathscr{C}=   \frac{I_{hot}}{I_{main}} = \frac{K_{hot}}{K_{main}} \frac{n}{n_0} \approx  \frac{K_{hot}}{K_{main}} \frac{n}{N_D} .    
 \label{C} 
 \end{equation}

\noindent
Since as shown above $n<<N_D$ and $p<<N_D$, one has $\delta n_0 << n_0$ because of electrical neutrality and the quantities  $\mathscr{B}$  and $\mathscr{C}$ are given by their approximate expressions in Eq. \ref{B} and Eq. \ref{C}, respectively. Thus the spatial profile of  $\mathscr{C}$ reveals that of the photoelectron concentration, while that of  $\mathscr{B}$    reflects the redistribution of the Fermi sea. The  spatial profiles of $\mathscr{C}$ and $\mathscr{B}$, normalized to unity at $z=0$, are shown in Curves c and d of  the various panels of Fig. \ref{Fig2}, respectively. The spatial charge redistribution obtained using  their analysis is summarized in  Fig. \ref{Fig4}.\

Immediately apparent is the fact that, in phase II, $\mathscr{B}$ is constant and $\mathscr{C} \approx n $ exhibits the same profile as $I_{main} \approx p$. The  latter finding   implies that $n \approx p$ so that $\delta n_0 <<p$ and that redistribution of intrinsic electrons is negligible. Since $\delta n_0 <<p$, one has $\mathscr{B}/\mathscr{B}_0 \approx 1$, in agreement with the observed constant value of  the latter quantity.\

In phase I,  at the smallest excitation power,  $\mathscr{B}$ at $z=0$ is  comparable with its value of 1 in Phase II, and the profile of $\mathscr{C}$ coincides with that of $I_{main}$ down to $z=0$. Thus, in the same way as for Phase II, charge redistribution is negligible. This no longer true when the excitation power increases,   since $\mathscr{B}$ at $z=0$ becomes smaller than its unit value of phase II, implying  that, in agreeement with charge neutrality,  there  appears an accumulation of intrinsic electrons at the excitation spot ($\delta n_0 >0$), which increases with excitation power. From the value of  $\mathscr{B} \approx n/p$ at $z=0$, we estimate that $n/p$ takes values of  $\approx 0.7$, $0.5$ and $0.1$ for panels B, C, and D, respectively.\

In phase III, the constant value of $\mathscr{C}$ indicates that $n$ is spatially constant.  In the same conditions, $\mathscr{B}$ increases by about one order of magnitude, nearly independently on excitation power,  implying that redistribution of intrinsic electrons is significant  ($\delta n_0<0$). Using  again $\mathscr{B} \approx n/p$  one  finds that for   $z>  15$ $\mu$m, one has $n \approx  10 p $, relatively independently on excitation power.\

Finally, in phase IV, observed for  $z>  17$ $\mu$m in panels C and D of  Fig. \ref{Fig2},  $\mathscr{C}$ starts to decrease, revealing, as seen from Eq. \ref{C}, a decrease of $n$. \

\section{Discussion}

In this section, we qualitatively interpret the spatial profiles in order to outline the possible mechanisms for charge transport. Note that since the relative modification of the charge in the Fermi sea is negligible ($\delta n_0<< n_0$) the spatial profile of line M is mostly caused by photohole drift in the electric field $E$ given by Fig. \ref{Fig4}, with a typical profile  for a locally homogeneous field of the type $exp( -z/ L_d)$, where the drift length is $ L_d = \mu_h E\tau_h $. \

\subsubsection{Phase I  :   z $ < 2$ $\mu$m}

In this spatial range the outward electric field required by a static equilibrium between diffusive and drift currents value increases with $p$ and thus with excitation power, in agreement with the power dependence of charge redistribution in this spatial range.\
An effect, which is likely to  decrease the outward electric field is the self-trapping of carriers by selective carrier injection in the subsurface depletion region \cite{park2019}. This effect originates from the spatial inhomogeneities of the photovoltage along the NW axis, which at least partly counterbalances the electric field  caused by redistribution of the Fermi sea. Such effect should be significant for the present NW since the depletion region (thickness  90 nm) is  a significant fraction of the NW radius.\

\subsubsection{ Phase II :  2 $\mu$m $<z<$ 2-10 $\mu$m}

In this spatial range, the  observation of an  exponential slowly-decaying profile, up to $ 10$ $\mu$m   at low excitation power, suggests long range hole diffusion in the fluctuations. This result is remarkable and contradicts the known difficulty of holes to undergo tunnel processes because of their large effective mass   \cite{levanyuk1981}.  In this range, a decrease of  $p$ induces a decrease of $\eta$ as seen from Eq. \ref{eqnonlin}, which itself induces an decrease of electric field, which is expected to be slightly  smaller than at smaller distance. Since hole transport is mostly due to drift in the electric field, the results of Sec. A would predict a decrease of the  drift length, at variance with the observations.\ 
This contradiction can be resolved since, as discussed in [I], the energy-distribution of holes in the fluctuations is  relatively narrow, at a given nonzero kinetic energy in the fluctuations  \cite{levanyuk1981, arnaudov1977}. It is then proposed that above a given distance, this energy is increased because of the electric field so that holes can more easily undergo tunneling processes. This  results in an increase of the length $\delta$, inducing an increase of $E$ and therefore of the hole drift length, as observed.\

\subsubsection{Phase III 2-10 $\mu$m $<z<$ 17-20 $\mu$m}

At a distance between $2$  $\mu$m and $10$ $\mu$m depending on excitation power, there occurs an abrupt change of  transport regime :  there occurs a significant Fermi sea redistribution,  the profile is no  longer exponential and  the increased value of $\vec{\nabla} p$ reveals, as seen from Eq. \ref{champel2}, that the electric field is increased.\ 

We propose that this electric field is  able to increase the average hole kinetic energy so that hopping processes have a strongly increased length, corresponding to an increase of $\delta$. While analysis of the hole transport near the top of the fluctuations is beyond the scope of the present work, it is shown here that a model considering a quasi-ballistic transport between two  phonon emission processes with a characteristic time $\tau_{ph}$,   explains the spatial profiles. Assuming for simplicity that the electric field $E$ is spatially constant in this region,  the hole  spatial profile is of the form $p(z) \approx exp(-z/ v_h \tau_{ph})$ where $v_h$ is the hole velocity so that $v_h \tau_{ph}$ is the mean free path for phonon emission. For ballistic transport, accelerated by an electric field,  one has $v_h= \sqrt{2 q m^{*-1}_h  Ez}$. One finally obtains    
\begin{equation}    
 p(z) \approx exp(-\sqrt{z/\mathscr{L}}) 
 \label{ballistic} 
 \end{equation} \
\noindent
where $\mathscr{L} = 2 (q/m^{*}_h) E \tau_{ph}^2$.  As shown in Fig. \ref{Fig2}, the spatial profiles in this phase are well approximated by Eq. \ref{ballistic} in view of the approximations made. Here $\mathscr{L}$ is $0.5$ $\mu$m, $1.1$ $\mu$m, $1.1$ $\mu$m and $1.8 $ $\mu$m for Curves a to d and therefore slightly increases with excitation power. Using $\mathscr{L} \approx 1$ $\mu$m and $\tau_{ph} \approx 0.25$ ps \cite{chebira92},  one finds a physically reasonable value of the internal electric field, of the order of 0.1 V/$\mu$m. This value is smaller than the unscreened electric field near the donor ($E_D/a_0^* \approx 0.6 V/\mu$m).\

Note that the distance corresponding to the beginning of phase III decreases with increasing excitation power. This is because in Phase II,  the internal electric field increases with excitation power so that the limit kinetic energy for hole quasi ballistic transport is reached earlier.\  

\subsubsection{Phase IV :  $z > 17-20$ $\mu$m }
The maximum distance over which photoelectrons can be transported can be found at high excitation power in panels C and D of  Fig. \ref{Fig2}. At these powers, for  $z>  17 \mu$m,  $\mathscr{C}$ starts to decrease, revealing, as seen from Eq. \ref{C}, a decrease of $n$. It is concluded that photoelectrons can be transported over distances as large as 20 $\mu$m. Since no change of slope is apparent on the profile of $I_{main}$ and since $\mathscr{B}$ also decreases in this spatial range, this suggests a decrease of $-\delta n_0$. It is thus   proposed that the limit to this distance is the resulting change of local electric field.\

\section{Conclusion}

We have investigated the mechanisms of charge and spin transport in HVPE-grown, plasma-passivated GaAs NWs, at 6K and as a function of excitation power. These  NWs have a n-type doping level in the  low  $ 10^{17}$ cm$^{-3}$ range implying that they  are metallic with a tail of density of states below the unperturbed bottom of the conduction band. The excellent NW quality is assessed from the luminescence spectra, which exhibit several well-resolved lines,  due to hot electrons, band-to-band and acceptor recombination.  This exceptional quality reflects the material quality as well as the  efficiency of the plasma surface passivation which was used. We have used a microluminescence technique in order to investigate at 6K the intensity and polarization spectra as a function of distance to the excitation spot, as well the intensity and polarization spatial profiles for selected energies in the spectrum.  The main results are\

a) These NW are potentially good candidates for transport.  Independently on excitation power, photoholes are  transported over distances as large as $ 20$ $\mu$m, while the photoelectron effective diffusion length can be even larger than this value.\

% The photoelectron spin polarization is large, of the order of 30 $\%$ at the excitation spot and is 15 $\%$ at a distance from this spot of $ 20$ $\mu$m.\ 

b) For distances up to $20$ $\mu$m, there builds up a strong internal field because of  spatial redistribution of the Fermi sea, including excess of intrinsic electrons near the excitation spot, compensated by depletion of these electrons ata distance larger than 2-10 $\mu$m depending on excitation power.\

c) This internal electric field strongly  favors  elementary tunneling processes, so that photoholes undergo a drifting transport with an increased mobility. Since the electric field value is determined by balance between electron diffusive and drift currents, there occurs a self-adjustment of this field to a value  such that the photohole drift current is comparable with that without fluctuations. \ 

In summary, the  reason why  minority photoholes and photoelectrons  can be  transported over record distances  as large as $ 20$ $\mu$m is the buildup of very large internal electric fields because of space charge effects caused by the redistribution of the Fermi sea, and the resulting strong increase of carrier mobility. In view of the  results of [I] concerning spin transport, it is anticipated that such NW can have potential applications for  charge and spin transport. \

\acknowledgments 

{We are grateful to D. Grebenkov for advice for the resolution of the diffusion equation shown in the supplementary material. This work was supported by Région Auvergne Rhône-Alpes (Pack ambition recherche; Convention  17 011236 01- 61617,  CPERMMASYF) and LabExIMobS3 (ANR-10-LABX-16-01). It was also funded by the program Investissements d 'avenir of the French ANR agency, by the French governement IDEX-SITE initiative 16-IDEX-0001 (CAP20-25), the European Commission (Auvergne FEDER Funds).}
 
\bibliographystyle{apsrev}
%\begin{thebibliography}

%\bibliography{cadizref}

\end{document}